\documentclass[sort&compress]{elsarticle}
\usepackage{microtype}
\usepackage{mathtools,mathrsfs,amssymb}
\usepackage[colorlinks=true, linkcolor = blue, hyperfootnotes=false, citecolor = blue]{hyperref}
\usepackage[english]{babel}
\usepackage{bm}

\graphicspath{{./figures/}}
\newcommand{\body}{{\{\t x\}}}
\newcommand{\Body}{{\{\t X\}}}
\newcommand{\Partial}{\mathcal{D}}

\newcommand{\flip}[1]{\bar{#1}}
\newcommand{\cross}{\times}
\renewcommand\phi{\varphi}
\renewcommand\[{\begin{equation}}
\renewcommand\]{\end{equation}}
\renewcommand{\t}[1]{\mathbf{#1}}
\newcommand{\gt}[1]{\bm{#1}}

\newcommand{\dd}{\mathrm{d}}

\newcommand{\tp}{\otimes}
\newcommand{\tps}{\otimes^s\!}

\newcommand{\abs}[1]{\left\lvert #1 \right\rvert}
\newcommand{\norm}[1]{\left\lVert #1 \right\rVert}
\newcommand{\scalar}[1]{\left\langle #1 \right\rangle}
\DeclareMathOperator{\tr}{tr}

\newcommand{\revise}[1]{#1}
\graphicspath{{./figures/}}
\begin{document}
\title{Willis elasticity from microcontinuum field theories: Asymptotics, microstructure-property relationships, and cloaking}
\author{H.~Nassar}
\ead{nassarh@missouri.edu}
\author{P.~Brucks}
\address{Department of Mechanical and Aerospace Engineering, University of Missouri, Columbia, Missouri 65211, USA}
\begin{abstract}
Willis elasticity is an effective medium theory for linearly elastic composites that incorporates an unusual coupling between stress and velocity, as well as between momentum and strain. Interest in the theory peaked following the discovery that its formulation is invariant under curvilinear changes of coordinates and that, consequently, it can be used to inverse-design ``invisibility'' cloaks for elastodynamics. That said, the microstructure-property relationships in Willis elasticity are poorly understood and, in particular, the mechanics that underlie the coupling are largely unknown. Thus, no such cloaks were constructed.

Here, we put forward the idea that Willis elasticity is a particular microcontinuum field theory where the (generalized) micro-displacements have been eliminated in favor of the macroscopic displacement field as if by Schur completion. The field theory is special in that it features an inertial coupling between the micro- and macro-displacements that, upon completion, re-emerges as the coupling term in Willis elasticity. Concretely, we analyze an asymptotic regime where mechanical lattices exhibit a kinematic enrichment with a strong (leading-order) inertial coupling. We provide, in closed-form, the resulting microstructure-property relationships. As an application, and in light of the gained insights, we design an ``invisibility'' cloak resolved into Willis-elastic mechanical lattices.
\end{abstract}
\maketitle

\section{Introduction}
The peculiar theory of linear elasticity here referred to as ``Willis elasticity'' first appeared as an implicit byproduct of investigations carried by J. R. Willis in the early 1980's into the effective dynamic behavior of random composites~\cite{Willis1980a, Willis1980, Willis1981, Willis1985}. It was not until 1997, it seems, that Willis recognized that the effective constitutive relations he had derived were unusually coupled~\cite{Willis1997}. He wrote ``It should be noted that the perturbation expansion demonstrates that, inevitably, the mean stress and mean momentum density are both coupled linearly to mean strain and mean velocity''. Formally, instead of the usual\footnote{Notations: $\gt\sigma$ is Cauchy's stress tensor, $\t e$ is infinitesimal strain tensor, $\t v$ is particle velocity, $\t p$ is linear momentum density, $\gt\mu$ is elasticity tensor and $\rho$ is mass density. Tensors of all orders higher than 1 are in bold.}
\[
    \gt\sigma = \gt\mu\t e,\quad
    \t p = \rho\t v,
\]
random composites turn out to be governed, on average, by two equations
\[
    \gt\sigma = \gt\mu\t e + \t s\t v,\quad
    \t p = \gt\rho\t v - \t s^\dagger\t e,
\]
coupled by what is now called a ``Willis coupling'' tensor $\t s$ and its adjoint $\t s^\dagger$. In the two cases investigated by Willis in his work of 1997, namely weakly heterogeneous composites and dilute suspensions, he notes that the coupling vanishes in the limit of low frequencies: $\t s \to \t 0$ as $\omega \to 0$. Since then, many authors have re-derived the equations of Willis elasticity, for periodic composites in particular~\cite{Willis2011, Shuvalov2011, Nassar2015a, Muhlestein2016}, and analyzed various aspects of the theory (e.g., accuracy, uniqueness, symmetry, etc.). What remained lacking is an understanding of the microstructural origins of strong low-frequency Willis coupling.

The question became of interest since it was discovered by Milton, Briane and Willis~\cite{Milton2006} that Willis elasticity is form-invariant under curvilinear changes of coordinates. In particular, Willis elasticity models composite materials and metamaterials that can be used as building blocks of ``invisibility'' cloaks for elastic waves. In that setup, we are faced with the inverse problem: the desired Willis coupling is known; but the microstructure that would produce it in its effective response is not. Given the poor understanding of the microstructure-property relationships in Willis elasticity, no such cloaks were ever constructed. Meanwhile, other venues for cloaking were investigated for acoustic waves, flexure waves, shear waves, and general elastodynamics and elastostatics~\cite{Parnell2012,Parnell2012a,Norris2012a,Farhat2009,Farhat2009a,Farhat2012,Nassar2018a,Nassar2019,Nassar2020,Xu2020,Chen2021}.

Recently, it was demonstrated that laminates whose properties are modulated in time in a periodic progressive fashion obey a form of Willis elasticity that is non-reciprocal (i.e., does not obey Maxwell-Betti reciprocity). The time modulation induces a bias in propagated frequencies that is similar to a Doppler effect and whose magnitude is proportional to the depth and frequency of the modulation. In the constitutive equations, the bias takes the form of a strong Willis coupling~\cite{Nassar2017}. This version of Willis elasticity, being non-reciprocal, is quite different from the one originally proposed by Willis and is not useful for cloaking applications. In other contributions, Willis coupling was interpreted as a coupling between the monopole and dipole moments of a scatterer or an interface~\cite{Melnikov2019}; such interpretations are not of concern here.

More relevant to our purposes is a paper by Milton~\cite{Milton2007a} where a mechanical lattice is shown to exhibit a strong Willis coupling at subwavelength scales. Milton's lattice features a resonator with a large mass that ensures the relevance of the resonator's dynamics at the macroscopic scale. But the required mass is so large that the lattice would be infinitely heavy. Milton's clever solution to this setback is to introduce a second resonator with an equally large but negative mass. In this fashion the effective mass density remains finite. The drawback to this design is that balancing large opposite masses means that the effective response is valid in the immediate vicinity of one specific frequency. Another drawback\footnote{This could also be a feature: it means that Milton's lattice is easily polarizable.} is the likely presence of strong boundary layers in any finite sample of the lattice since the delicate balance of positive and negative masses might not hold near boundaries. In another development, Boutin, Auriault and Bonnet~\cite{Boutin2018} carried out a systematic analysis of the subwavelength asymptotic behavior of high-contrast two-phase composites with one connected phase, namely the matrix (properties $\gt\mu^m,\rho^m$), and one disconnected phase, namely the inclusion (properties $\gt\mu^i, \rho^i$). They found that for an asymptotic scaling that favors resonance of the form
\[
    \gt\mu^i \sim \epsilon \gt\mu^m,\quad \rho^i \sim \rho^m/\epsilon,
\]
with $\epsilon$ being the factor of separation of scales, the composite behaves, to leading order, like a Willis medium but only at frequencies where the effective mass of the inclusion vanishes. Fittingly, they called this phenomenon ``anti-resonance''. The ``anti-resonant'' composite is basically the continuum version of Milton's lattice of 2007 and possesses the same virtues and drawbacks. See also~\cite{Qu2022} for variations on Milton's design.

In the following, mechanical lattices whose effective medium theory is Willis elasticity are referred to as \emph{Willis-elastic}. The main purpose of the paper is to propose a novel paradigm for the design of Willis-elastic mechanical lattices that is more powerful than that based on resonance and anti-resonance. In particular, the resulting Willis coupling will survive over an uninterrupted broad spectrum of low frequencies. The proposal proceeds in two steps. In step 1, we demonstrate how Willis elasticity can be obtained from a particular class of microcontinuum field theories provided two conditions: $(i)$ the theory is kinematically enriched, and $(ii)$ the kinematic enrichment is inertially coupled to the macroscopic displacement. By elimination of the micro-displacements in favor of the macroscopic displacement field, as if by Schur completion, the inertial coupling re-emerges as a Willis coupling. In step 2, we investigate mechanical lattices whose effective medium theory satisfies conditions $(i)$ and $(ii)$. Condition $(i)$ is satisfied in mechanical lattices with slightly misaligned bonds, the twisting modes of the bonds being in correspondence with the degrees of kinematic enrichment~\cite{Nassar2020c}. Condition $(ii)$ is easier to satisfy: it suffices to offset the center of mass of the unit cell. These lattices are then proven, by means of a leading-order asymptotic analysis, to exhibit the desired effective medium theory. As an application, we investigate a particular case of Willis-elastic materials that appear in 3D transformation-based cloaking. The desired Willis-elastic materials are then resolved into Willis-elastic mechanical lattices. A numerical demonstration is presented in 2D in a case where the transformation is conformal. The main finding of the paper, namely that Willis elasticity is a particular micro-continuum field theory, is disappointing in a sense: it implies that \emph{Willis elasticity does not describe any new wave phenomena that are not already within the reach of generalized elasticity.} That said, Willis elasticity proves useful to perform some inverse design tasks and the application to cloaking goes to prove that usefulness.

\section{Step 1: Willis elasticity from microcontinuum field theories}
\begin{figure}
    \centering
    \includegraphics[width=\linewidth]{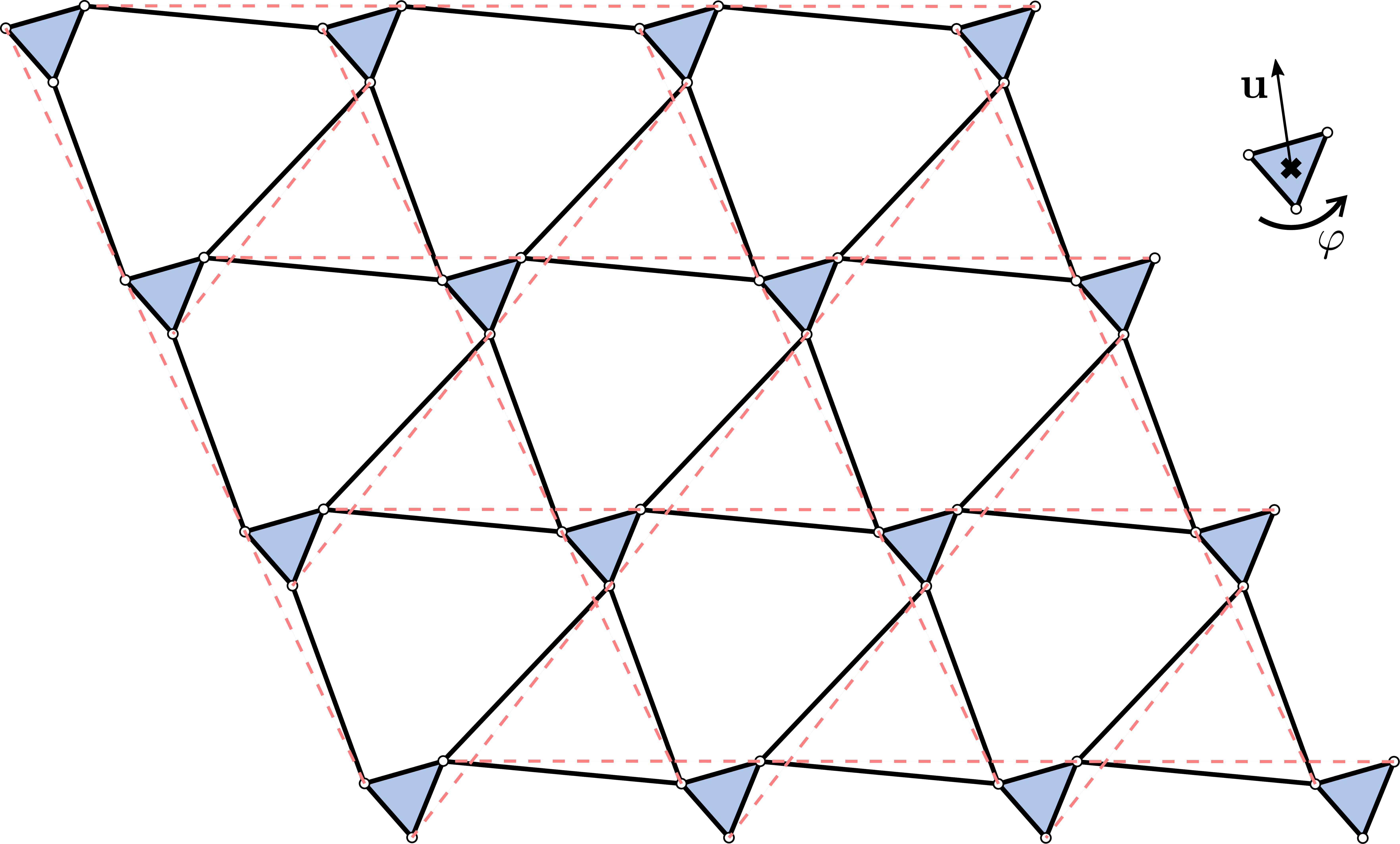}
    \caption{A Kagome lattice with a slight misalignment between the elastic bonds (solid segments) and the lattice directions (dashed lines). The small circles are pin joints. Inset shows the macroscopic degrees of freedom.}
    \label{fig:kagome}
\end{figure}
\subsection{The main idea}
The inspiration comes from a recent contribution to the continuum modeling of 2D Kagome lattices~\cite{Nassar2020c}. Therein, it is shown that Kagome lattices, with slightly misaligned fibers, behave according to a microcontinuum field theory called ``microtwist elasticity'' (Fig.~\ref{fig:kagome}). The constitutive and balance equations of the theory are
\[\label{eq:microtwist}
\begin{bmatrix}
\gt\sigma\\
\gt\xi\\
-\xi
\end{bmatrix} = 
\begin{bmatrix}
    \gt\mu & \t b & \t m\\
    \t b & \t d & \t a\\
    \t m & \t a & \eta
\end{bmatrix}
\begin{bmatrix}
    \t e \\ \gt\nabla\phi \\ \phi
\end{bmatrix},\qquad
\left\{
\begin{aligned}
\gt\nabla\gt\sigma &= -\omega^2(\rho\t u + \rho\phi\cross\t c), \\
\gt\nabla\gt\xi + \xi &= -\omega^2(\rho\t c\cross\t u + j\phi).
\end{aligned}\right.
\]
As for notations: $\gt\sigma$, $\gt\xi$ and $\xi$ are stress measures dual to the infinitesimal strain tensor $\t e$, the gradient of micro-rotation $\gt\nabla\phi$ and the micro-rotation $\phi$; $\gt\mu$ through $\eta$ are fully symmetric constitutive tensors of various orders; $\rho$ is mass density, $j$ is the moment of micro-inertia and $\t c$ is an inertial coupling term to be discussed hereafter; $\t u$ is the macroscopic displacement field and $\phi$ is a field of micro-rotations; last, $\omega$ is angular frequency. Also, whenever two tensors meet in a monomial, they are maximally contracted, e.g, $\gt\nabla\gt\sigma$ has components $\sigma_{ij,i}$. The cross product $\cross$ is defined in 2D by identifying scalars with vectors normal to the 2D plane.

Dismissing specifics, these equations are typical of enriched microcontinuum field theories of the first gradient minus one detail: the inertial coupling $\t c$ is usually set to zero~\cite{Mindlin1964, Germain1973, CemalEringen1999}. The main claim is that the inertial coupling $\t c$ is the micro-continuum manifestation of Willis coupling. Letting go of Kagome lattices and focusing on the field equations, suppose that the constitutive relations are singular in a way that systematically produces zero hyperstress $\gt\xi$. In other words suppose $\t b$, $\t d$ and $\t a$ vanish. Then, the model simplifies greatly into
\[
\begin{bmatrix}
\gt\sigma\\
-\xi
\end{bmatrix} = 
\begin{bmatrix}
    \gt\mu & \t m\\
    \t m & \eta
\end{bmatrix}
\begin{bmatrix}
    \t e \\ \phi
\end{bmatrix},\qquad
\left\{
\begin{aligned}
\gt\nabla\gt\sigma &= -\omega^2(\rho\t u + \rho\phi\cross\t c), \\
\xi &= -\omega^2(\rho\t c\cross\t u + j\phi).
\end{aligned}\right.
\]
Thus, the second field equation becomes algebraic and it can be solved for $\phi$ which then can be eliminated from the other equations in favor of $\t u$ and $\t e$. Namely,
\[
    \phi = \frac{-\t m\t e + \omega^2 \rho\t c\cross\t u}{\eta-\omega^2j}
\]
implies
\[
\left.
\begin{aligned}
    \gt\sigma &= \gt\mu^*\t e + \omega^2\t s\t u,\\
    \gt\nabla\gt\sigma &= -\omega^2(\gt\rho^*\t u - \t s^\dagger\t e),
\end{aligned}\right\}
\quad \text{with} \quad \left\{
\begin{aligned}
\gt\mu^* &\equiv \gt\mu - \frac{\t m\tp\t m}{\eta-\omega^2j},\\
\t s &\equiv \rho\frac{\t m\tp\flip{\t c}}{\eta-\omega^2j},\\
\gt\rho^* &\equiv \rho\left(\t I + \omega^2\rho\frac{\flip{\t c}\tp\flip{\t c}}{\eta-\omega^2j}\right),
\end{aligned}
\right.
\]
where $\t I$ is the identity and $\flip{\t c}$ is $\t c$ rotated through $\pi/2$. Upon introducing an apparent linear momentum density $\t p\equiv -i\omega(\gt\rho^*\t u - \t s^\dagger \t e)$, it comes that the above equations describe a Willis-elastic continuum with Willis coupling $i\omega\t s$.

\subsection{Discussion}
The conclusion of the foregoing derivation is that \emph{a microcontinuum field theory with a kinematic enrichment that is coupled to both stress and momentum is equivalent to a Willis theory}. Now it is not unusual for the enrichment to contribute to stress, e.g., through tensor $\t m$. However, it is the dominant view, in microcontinuum field theories~\cite{Mindlin1964, Germain1973, CemalEringen1999}, that the inertial coupling $\t c$ is $\t 0$. The justification for disregarding $\t c$ is that it is always possible to choose the origin of micro-displacements (e.g., the center of rotation) as the center of mass of the micro-body. Indeed, it will become clear in the following section that $\t c$ is exactly that: the position of the center of mass within a unit cell. Should $\t c$ be non-zero, it is possible to shift the origin to make it so. That said, it should be recognized that \emph{shifting the origin of micro-displacements influences the elastic constitutive properties}. Using the example of equations~\eqref{eq:microtwist}, one can eliminate $\t c$ with the change of variable 
\[
    \t u \mapsto \t u - \phi\cross \t c
\]
which amounts to shifting the center of micro-rotations to match it with the center of mass. Then, $\t e\mapsto \t e - \flip{\t c}\tps\gt\phi$ and it follows that
\[
\t d \mapsto \t d + \flip{\t c}\gt\mu\flip{\t c} - 2\t b\flip{\t c},\quad
\t b \mapsto \t b - \gt\mu\flip{\t c},\quad
\t a \mapsto \t a - \t m\flip{\t c}.
\]
Changing variables, and the constitutive law, in such a fashion provides another valid description of the same field theory. But, in the case where the constitutive law is degenerate with $\t b = \t 0$, $\t d = \t 0$ and $\t a = \t 0$, it is far more convenient to deal with the consequences of $\t c\neq\t 0$ than to change variables and deal with the consequences of $\t b\neq \t 0$, $\t d\neq \t 0$ and $\t a\neq \t 0$.

In cases where the constitutive law is not degenerate ($\t b\neq \t 0$ and so on), it is still possible to ``forcefully'' eliminate $\phi$ in favor of $\t u$ and $\t e$ assuming, say, homogeneous boundary conditions (e.g., $\phi=0$). The corresponding expression of $\phi$ will involve non-local integral operators. The resulting Willis theory will feature constitutive relations that are non-local in space with non-decaying boundary-dependent kernels as in the original formulation by Willis~\cite{Willis2011}. But it is difficult to imagine a scenario where such a formulation would be more convenient than the full local field theory with kinematic enrichment.

\section{Step 2: Asymptotics and microstructure-property relationships}
Our focus shifts to the study of some mechanical lattices whose effective medium theory is a microcontinuum field theory of the form described in the previous section, i.e., degenerate with an inertially-coupled kinematic enrichment. The effective medium theory is derived asymptotically to leading order in the size of a unit cell. The mechanical lattices to be analyzed are composed of several families of parallel, almost straight, fibers. Again, this choice is motivated by a recent study of 2D Kagome lattices~\cite{Nassar2020c}. Therein, it is shown that a slight misalignment in the fibers relaxes the coupling between fiber twisting and stretching and effectively turns twisting degrees of freedom into a kinematic enrichment. Details follow starting with the case of a single fiber.

\subsection{One fiber}
\begin{figure}
    \centering
    \includegraphics[width=\linewidth]{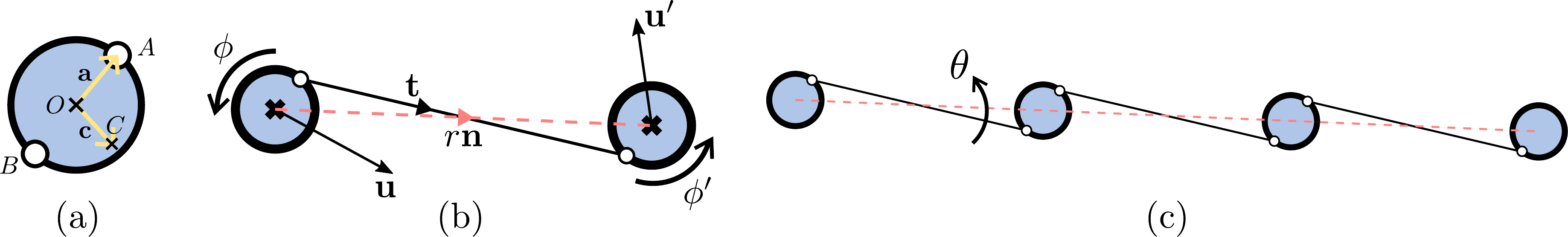}
    \caption{A fiber, annotated: (a) the rigid body; (b) two interacting bodies; (c) the whole periodic fiber embedded in 2D or 3D. Small circles are pin joints; solid lines are elastic bonds and dashed lines are lattice directions.}
    \label{fig:onefiber}
\end{figure}

We call a ``fiber'' a 1D chain of springs and rigid bodies embedded in 3D space (Fig.~\ref{fig:onefiber}). The bodies are identical. Each has a mass $m$ and a moment of inertia $\t J$ relative to some point $O$. The center of mass is at a point $C$ with $\t c\equiv OC$. The spring constant is $k$ and the spring is attached at two points $A$ and $B$ such that $\t a \equiv OA$ and $\t b \equiv OB$ and is oriented parallel to unit vector~$\t t$. It is enough for our purposes to consider the case $\t b = - \t a$. Let $\t u$ be the displacement at point $O$ and $\gt\phi$ be the infinitesimal rotation vector about point $O$. Then, the next body is displaced through $\t u' = \t u + r\partial\t u$ and rotated through $\gt\phi' = \gt\phi + r\partial\gt\phi$, with $r$ being the spacing between two consecutive bodies and $\partial$ denoting a derivative in the direction of the chain, namely, $\partial \equiv \t n\gt\nabla$ where $r\t n$ is the lattice vector. Therein, first-order Taylor expansions relative to $r$ are used under the assumption that the displacements of the bodies derive from smooth functions $\t u = \t u(\t x)$ and $\gt\phi = \gt\phi(\t x)$ where $\t x$ is body position. Then, the elongation of a spring takes the form
\[
\scalar{\t u'- \gt\phi'\cross\t a - \t u - \gt\phi\cross\t a,\t t} = \scalar{r\partial\t u -2 \gt\phi\cross\t a - r\partial\gt\phi\cross\t a, \t t},
\]
and the contribution of one unit cell to the Lagrangian is
\[
L = \frac{1}{2}k\scalar{r\partial\t u -2 \gt\phi\cross\t a - r\partial\gt\phi\cross\t a, \t t}^2  - \frac{1}{2}m\scalar{\dot{\t u},\dot{\t u}} - m\scalar{\dot{\t u}, \dot{\gt\phi}\cross\t c} - \frac{1}{2}\dot{\gt\phi}\t J\dot{\gt\phi},
\]
where the brackets denote the dot product. Let $\ell\equiv L/V$ be the Lagrangian density where $V$ is, for now, a nominal volume of order $O(r^3)$.

The asymptotic regime of interest is one where $\gt\phi$ plays the role of a kinematic enrichment. For this to happen, displacements due to $\gt\phi$ should be comparable to $\t u$, namely: $\t u = O(r\gt\phi)$. At the same time, elongations $\scalar{r\partial\t u,\t t}$ should be comparable to $\scalar{\gt\phi\cross\t a,\t t}$. In general, the former, of order $O(r^2\gt\phi)$, is dominated by the latter, of order $O(r\gt\phi)$ because $\t a=O(r)$. There is an exception however: if $\t a$ happens to align with $\t t$, then $\scalar{r\partial\t u,\t t}$ dominates $\scalar{\gt\phi\cross\t a,\t t}=0$. Hence, in cases where $\t a$ and $\t t$ are slightly misaligned, the two contributions to elongations could balance. Note that this also means that $\t t$ and $\t n$ are slightly misaligned and that the last contribution to elongations, namely $\scalar{r\partial\gt\phi\cross\t a,\t t}=O(r^3\gt\phi)$, can be neglected.

In summary, let $\t t = \t n + O(r)$ and let $\gt\theta\equiv\t t\cross\t n = O(r)$ be the small angular misalignment so that $r\gt\theta=2\t t\cross\t a=O(r^2)$ (see Fig.~\ref{fig:onefiber}c). Furthermore, let $m\equiv\rho V$ derive from a mass density $\rho=O(1)$; let $\t J=\t j V$ derive from a moment of inertia density $\t j = O(r^2)$; and, let $k\equiv EV/r^2=O(r)$ derive from a Young's modulus $E=O(1)$. Then, to leading order, namely $O(\abs{\t u}^2)$, the Lagrangian density reads
\[
\ell = \frac{1}{2}E\left(\scalar{\partial\t u ,\t n}+ \scalar{\gt\phi, \gt\theta}\right)^2  - \frac{1}{2}\rho\scalar{\dot{\t u},\dot{\t u}} - \rho\scalar{\dot{\t u}, \dot{\gt\phi}\cross\t c} - \frac{1}{2}\dot{\gt\phi}\t j\dot{\gt\phi}.
\]

Consider one last addition to the architecture: embed the fiber in a soft, light, elastic matrix. The interaction between the matrix and the rigid bodies involves contributions quadratic in $(r\partial\t u, r\gt\phi,r^2\partial\gt\phi)$. Given that $\t u=O(r\gt\phi)$, only the term in $\gt\phi$ survives. Thus, the Lagrangian density becomes
\[
\ell = \frac{1}{2}E\left(\scalar{\partial\t u ,\t n}+\scalar{\gt\phi, \gt\theta}\right)^2 + \frac{1}{2}\gt\phi\t K\gt\phi - \frac{1}{2}\rho\scalar{\dot{\t u},\dot{\t u}} - \rho\scalar{\dot{\t u}, \dot{\gt\phi}\cross\t c} - \frac{1}{2}\dot{\gt\phi}\t j\dot{\gt\phi},
\]
where $\t K=O(r^2)$ is a symmetric positive definite tensor characterizing the interaction between the rigid bodies and the soft matrix. Another interpretation of $\t K$ is that it is the stiffness matrix of torsional springs linking the rigid bodies to the bonds.
\subsection{Multiple fibers}
\begin{figure}
    \centering
    \includegraphics[width=\linewidth]{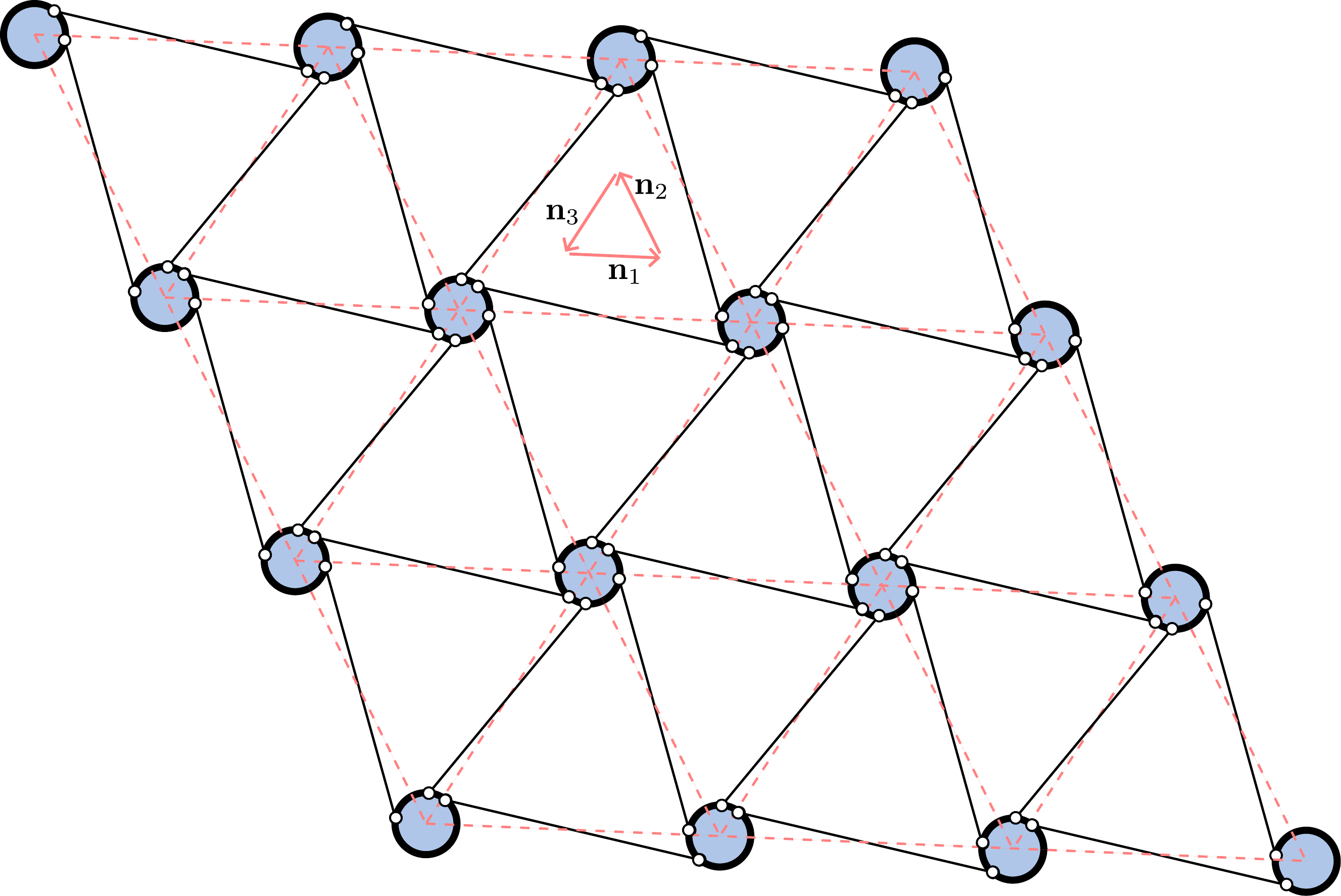}
    \caption{Multiple fibers.}
    \label{fig:multifibers}
\end{figure}

Now consider copies of the same spring-mass chain all parallel and equally spaced; then consider multiple families of chains, indexed with $i=1\dots I$, each characterized by its unit director $\t n^i$, Young's modulus $E^i$, and its misalignment vector $\gt\theta^i$ (Fig.~\ref{fig:multifibers}). Then, the total Lagrangian density is
\[
\ell = \frac{1}{2}\sum_i E^i\left(\scalar{\partial_i\t u ,\t n^i}+ \scalar{\gt\phi, \gt\theta^i}\right)^2 + \frac{1}{2}\gt\phi\t K\gt\phi  - \frac{1}{2}\rho\scalar{\dot{\t u},\dot{\t u}} - \rho\scalar{\dot{\t u}, \dot{\gt\phi}\cross\t c} - \frac{1}{2}\dot{\gt\phi}\t j\dot{\gt\phi},
\]
with $\partial_i\equiv \t n^i\gt\nabla$.

Therefore, to leading order, the architected mechanical lattice behaves according to a microcontinuum field theory that is $(i)$ kinematically enriched with a micro-rotation vector $\gt\phi$; $(ii)$ degenerate since it does not feature the micro-rotation gradient; and, $(iii)$ inertially coupled with coupling $\t c$.

A few comments are in order:
\begin{enumerate}
    \item This particular microcontinuum field theory was named ``microtwist elasticity'' in the context of 2D and 3D Kagome lattices. It features a kinematic enrichement identified as a micro-rotation but is different from Cosserat elasticity. In particular, it is clear that $\ell$ only depends on the symmetric part of strain through $\scalar{\partial_i\t u,\t n^i}=\t n^i\t e\t n^i$ with $\t e\equiv\gt\nabla\tps\t u$. The corresponding Cauchy's stress is symmetric.
    \item As brought up earlier, it is possible to eliminate the inertial coupling $\t c$ by considering the center of mass $C$, not $O$, to be the center of rotation. This however will bring back the dependence of $\ell$ over $\gt\nabla\gt\phi$. Point $O$ is special because it is the intersection point of all fibers, asymptotically speaking.
    \item Should the misalignment be large, micro-rotations would dominate the elastic response which, paradoxically, would impede them.\footnote{This is similar to how certain shells respond in pure bending because stretching is dominates the elastic energy and is too expensive to appear.} Ultimately, each degree of kinematic enrichment corresponds to, on the lattice level, a zero-energy deformation mode or to a near-zero-energy deformation mode.
    \item Several earlier contributions have investigated lattices such as the one depicted in Fig.~\ref{fig:multifibers} (see, e.g.,~\cite{Spadoni2009,Bacigalupo2014,Rosi2016}) but only in cases where the misalignment is large and, accordingly, where microstructural effects are only substantial at high frequencies.
\end{enumerate}

\subsection{Willis elasticity}
The constitutive and balance equations of the field theory are straightforward to obtain. They read
\[
\begin{bmatrix}
\gt\sigma\\
-\gt\xi
\end{bmatrix} = 
\begin{bmatrix}
    \gt\mu & \t m^\dagger\\
    \t m & \gt\eta
\end{bmatrix}
\begin{bmatrix}
    \t e \\ \gt\phi
\end{bmatrix},\qquad
\left\{
\begin{aligned}
\gt\nabla\gt\sigma &= \rho\ddot{\t u} + \rho\ddot{\gt\phi}\cross\t c, \\
\gt\xi &= \rho\t c\cross\ddot{\t u} + \t j\ddot{\gt\phi},
\end{aligned}\right.
\]
with
\[
\gt \eta = \t K + \sum_i E^i\gt\theta^i\tp\gt\theta^i,\quad
\t m = \sum_i E^i\gt\theta^i\tp\t n^i\tp\t n^i,\quad
\gt\mu = \sum_i E^i\t n^i\tp\t n^i\tp\t n^i\tp\t n^i.
\]

Note that the second field equation is an ordinary differential equation in time and can be solved for $\gt\phi$ provided initial conditions. This leads to a Willis elasticity theory with a history-dependent behavior. Equivalently, we consider a steady state of angular frequency $\omega$ and obtain
\[
\gt\phi = \left(\gt\eta- \omega^2\t j\right)^{-1}\left(-\t m\t e+\omega^2\rho\t c\cross\t u\right).
\]
Substituting back into the first field equation provides the equations of Willis elasticity
\[
\left.
\begin{aligned}
    \gt\sigma &= \gt\mu^*\t e + \omega^2\t s\t u,\\
    \gt\nabla\gt\sigma &= -\omega^2(\gt\rho^*\t u - \t s^\dagger\t e),
\end{aligned}\right\}
\quad \text{with} \quad \left\{
\begin{aligned}
\gt\mu^* &\equiv \gt\mu - \t m^\dagger\cdot\left(\gt\eta- \omega^2\t j\right)^{-1}\cdot\t m,\\
\t s &\equiv \rho\t m^\dagger\cdot\left(\gt\eta- \omega^2\t j\right)^{-1}\cdot\flip{\t c},\\
\gt\rho^* &\equiv \rho\left[\t I - \omega^2\rho\flip{\t c}\cdot\left(\gt\eta- \omega^2\t j\right)^{-1}\cdot\flip{\t c}\right],
\end{aligned}
\right.
\]
where $\cdot$ contracts tensors over their closest two indices and $\flip{\t c}$ is a skew tensor whose axial vector is $\t c$.

\subsection{An example in 2D}
The above theory can be specified to 2D. Simply the tensors $\gt\phi$, $\gt\eta$, $\gt\theta^i$, $\t K$ and $\t j$ become the scalars $\phi$, $\eta$, $\theta^i$, $K$ and $j$. Let the fibers run in three directions $2\pi/3$ apart with equal misalignment $\theta^i=\theta$ and equal spring constants $k^i=k$. The natural scaling for the spring constants is $k=EA/r^2 = O(1)$ with $A=O(r^2)$ being the unit cell area. Then, the constitutive tensors of the full theory are
\[
    \eta = K + 3E\theta^2,\quad
    m_{ij} = \frac{3}{2}\theta E\delta_{ij}, \quad
    \mu_{ijkl} = \frac{3}{8}E(\delta_{ij}\delta_{kl}+\delta_{ik}\delta_{jl}+\delta_{il}\delta_{jk}).
\]
The expression of $\phi$ simplifies into
\[
\phi = \frac{-3\theta E \tr(\t e)/2 + \omega^2\rho\t c\cross\t u}{\eta - \omega^2 j}.
\]
Last, the reduced constitutive tensors are
\[
    \begin{split}
    \mu^*_{ijkl} &= \frac{3}{8}E(\delta_{ij}\delta_{kl}+\delta_{ik}\delta_{jl}+\delta_{il}\delta_{jk}) - \frac{9}{4}\frac{\theta^2 E^2}{\eta-\omega^2 j}\delta_{ij}\delta_{kl} \\
    s_{ijk} &= \frac{3}{2}\rho\frac{\theta E}{\eta-\omega^2 j}\delta_{ij}\flip{c}_k 
    \\
    \rho^*_{ij} &= \rho\left(\delta_{ij} + \omega^2\rho\frac{\flip{c}_i\flip{c}_j}{\eta-\omega^2 j}\right).
    \end{split}
\]
The reduced effective properties are dispersive because of the underlying dynamics of $\phi$. The misalignment brings the dispersion to the bulk modulus of $\gt\mu^*$ and the offset of the center of mass brings it to mass density. Close to the resonance frequency $\omega=\sqrt{\eta/j}$, both bulk modulus and mass density (in direction $\flip{\t c}$) become negative. These effects are not due to the Willis coupling per se, but both are needed for the coupling to survive. The misalignment and the offset of the center of mass are both of order $O(r)$ but the moment of inertia density is of order $O(r^2)$ making the Willis coupling of order $O(1)$. The offset also has the side effect of making the reduced mass density anisotropic.

\subsection{Back to a single fiber}
We return to the case of a single fiber embedded in 3D, then in 2D, in anticipation of the next section where we design an ``invisibility'' cloak for elastodynamics. Suppose that $\gt\eta-\omega^2\t j$ is an isotropic tensor identifiable with a scalar $\eta-\omega^2 j$. Then the Willis constitutive tensors are
\[\label{eq:1fiber3d}
\begin{split}
\gt\mu^* &= E^*\t n\tp\t n\tp\t n\tp\t n,\qquad\qquad E^* \equiv E - \frac{E^2\norm{\gt\theta}^2}{\eta-\omega^2 j},\\
\t s &= \frac{\rho E}{\eta-\omega^2 j} \t n\tp \t n\tp (\gt\theta\cross\t c),\\
\gt\rho^* &= \rho\t I - \frac{\rho^2\omega^2}{\eta-\omega^2 j}\flip{\t c}\tp\flip{\t c}.
\end{split}
\]
In 2D, these become
\[\label{eq:1fiber2d}
\begin{split}
\gt\mu^* &= E^*\t n\tp\t n\tp\t n\tp\t n,\quad\text{with}\quad E^* \equiv E - \frac{E^2\theta^2}{\eta-\omega^2 j},\\
\t s &= \frac{\rho E \theta}{\eta-\omega^2 j} \t n\tp \t n\tp \flip{\t c},\\
\gt\rho^* &= \rho\t I - \frac{\rho^2\omega^2}{\eta-\omega^2 j}\flip{\t c}\tp\flip{\t c},
\end{split}
\]
where $\flip{\t c}$ can be interpreted again as $\t c$ rotated through $\pi/2$.

\section{Application: Willis-elastic mechanical lattices for cloaking}
\begin{figure}
    \centering
    \includegraphics[width=\linewidth]{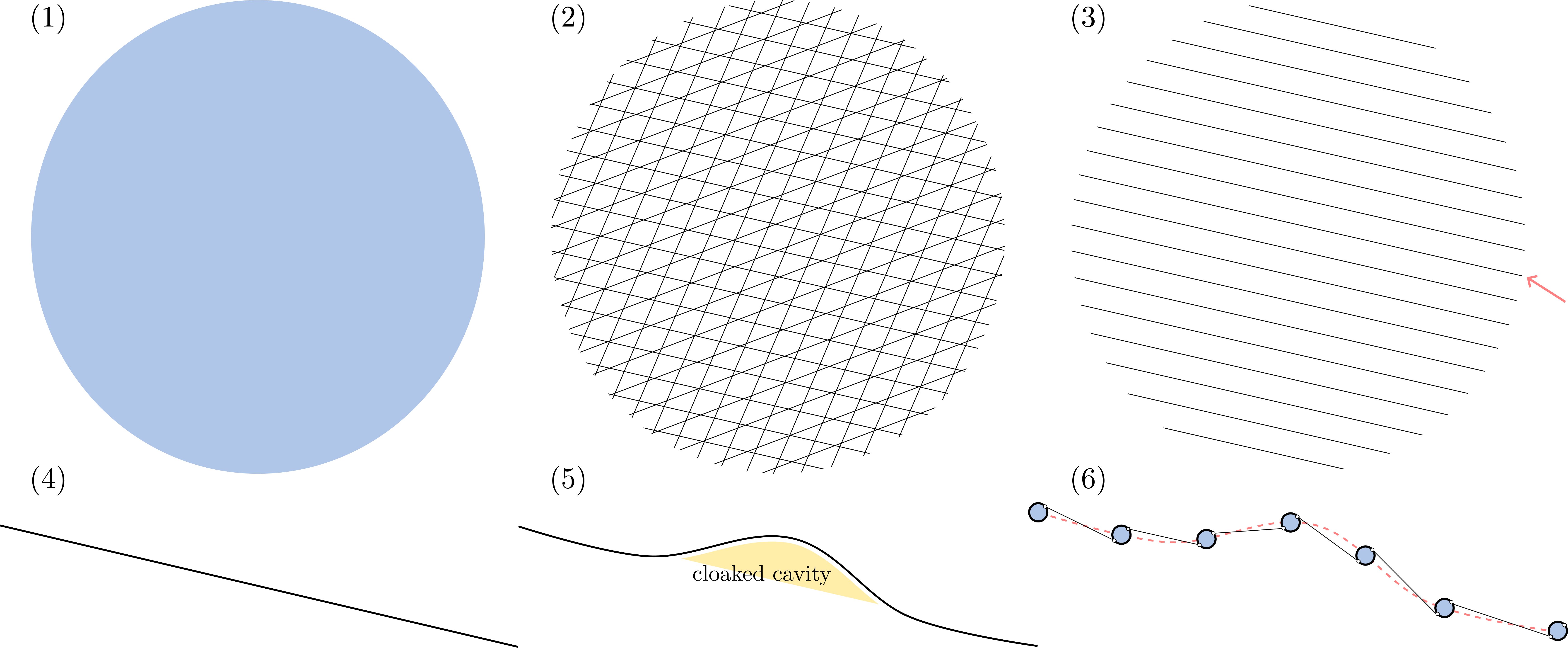}
    \caption{Design methodology: start with a reference medium (1) whose elasticity tensor is that of a set of elastic fibers (2); extract one family of fibers (3) and then one specific fiber (4); apply the transformation method to the specific fiber (5) and resolve the resulting medium into an elastic lattice (6).}
    \label{fig:design}
\end{figure}

Consider an inclusion of arbitrary shape embedded in an infinite homogeneous medium of elasticity $\gt\mu_o$ and mass density $\rho_o$. An ``invisibility'' cloak is a coating of the inclusion that eliminates any scattering off of the now-coated inclusion regardless of the properties of the inclusion. The properties of the cloak can be obtained using the transformation method and turn out to be those of a particular Willis-elastic medium. Here, we resolve said Willis-elastic medium into Willis-elastic mechanical lattices. Computations are carried for a general transformation in 3D then specified to a conformal transformation in 2D. We conclude with a numerical demonstration of cloaking where the simulation is performed at the level of the lattice. This should be the first demonstration of cloaking of that kind. Throughout, we restrict attention to the case where the elasticity tensor of the background medium is of the form
\[
\gt\mu_o = E_o \t N\tp\t N\tp\t N\tp\t N.
\]
The background medium then appears to be composed of a single family of straight parallel fibers running in direction $\t N$. The results generalize immediately to cases with
\[
\gt\mu_o = \sum_i E^i_o \t N^i\tp\t N^i\tp\t N^i\tp\t N^i,
\]
by superposition. An illustration of the design methodology is presented in Fig.~\ref{fig:design}. For more details on the transformation method, see~\cite{Norris2008,Norris2011,Norris2015}.

\subsection{The 3D transformation method for a single fiber}
Consider an elastic medium $\Body\subset\mathbb{R}^3$ with a Lagrangian density of the form
\[
\mathcal{L} = \frac{1}{2}E_o\scalar{\Partial\t U,\t N}^2 - \frac{\omega_o^2}{2}\rho_o\scalar{\t U,\t U}.
\]
The elastic medium behaves as a single family of straight parallel non-interacting fibers that run in the direction of the unit vector $\t N$. Only strain in direction $\t N$ is relevant: $\Partial\equiv \t N\gt\nabla$. Perform the change of variables
\[
    \t x = \gt\psi(\t X),\quad \dd\t x = \t F\dd \t X,\quad \t U(\t X) = \t F^\dagger(\t X)\t u(\t x), \quad 
\]
then the Lagrangian becomes
\[
\int_\Body \mathcal{L} = \int_\body
\frac{1}{2}\frac{E_o}{\abs{\t F}}\scalar{\partial(\t F^\dagger\t u),\t N}^2
-\frac{1}{2}\frac{\rho_o}{\abs{\t F}}\scalar{\t F^\dagger\t u,\t F^\dagger\t u},
\]
where
\[
\body = \gt\psi(\Body),\quad
\partial \equiv \t n\gt\nabla,\quad 
\t n \equiv \t F\t N,\quad
\abs{\t F} \equiv \det \t F.
\]
Further expansion leads to
\[
\int_\Body \mathcal{L} = \int_\body
\frac{1}{2}\frac{E_o}{\abs{\t F}}\left(\scalar{\partial\t u,\t n}+\scalar{\t u,\partial\t n}\right)^2
-\frac{\omega_o^2}{2}\scalar{\t u,\gt\rho \t u} \equiv \int_\body\ell,
\]
with $\gt\rho \equiv \rho_o\t F\t F^\dagger/\abs{\t F}$.

In the paradigm of the transformation method, domain $\body$ is identified with a physical body of Lagrangian density $\ell$ in which case bodies $\Body$ and $\body$ would be indistinguishable for their displacement fields are in a one-to-one correspondence. The challenge that remains is to find a medium, or rather an effective medium, whose motion is governed by the identified density $\ell$. This medium $\body$ will be referred to as the ``cloak''.

The insight provided by Milton, Briane and Willis~\cite{Milton2006} is that $\ell$ describes a Willis-elastic medium. In the present particular case, the Willis constitutive tensors deduced form $\ell$ are
\[
    \gt\mu^* = \frac{E_o}{\abs{\t F}} \t n\tp\t n\tp\t n\tp\t n,\quad
    \omega_o^2\t s = \frac{E_o}{\abs{\t F}} \t n\tp\t n\tp \partial\t n,\quad
    \omega_o^2\gt\rho^* = \omega_o^2\gt\rho - \frac{E_o}{\abs{\t F}}\partial\t n\tp\partial\t n.
\]
It is clear that, except for mass density $\gt\rho^*$, these constitutive tensors are in the span of the effective tensors found by homogenization in equation~\eqref{eq:1fiber3d} for a single fiber of springs and rigid bodies. As for mass density $\gt\rho^*$, it is always possible to achieve using a suitable resonator embedded in each rigid body (see, e.g., Milton and Willis~\cite{Milton2007}).

\subsection{Conformal transformations in 2D}
In 2D, let the transformation $\gt\psi$ be conformal so that $\t F = \lambda\t R$ is a rotation $\t F$ composed with a stretch of factor $\lambda$. Then $\abs{\t F}=\lambda^2$ and $\t F\t F^\dagger=\lambda^2\t I$. Hence, $\gt\rho=\rho_o\t I$. It will also prove convenient to redefine $\t n$ so that it is unitary. Thus, let $\t n\equiv\t R\t N$. All in all, the Willis-elastic materials needed for 2D conformal cloaking are
\[\label{eq:cloakProperties}
\begin{split}
\gt\mu^* &= E^* \t n\tp\t n\tp\t n\tp\t n,\quad\text{with}\quad E^* = E_o\lambda^2,
\\
\omega_o^2\t s &= \lambda E_o \t n\tp\t n\tp \partial(\lambda\t n),\\
\omega_o^2\gt\rho^* &= \omega_o^2\rho_o\t I - E_o\partial(\lambda\t n)\tp\partial(\lambda\t n).
\end{split}
\]
Remarkably, all three constitutive tensors are in direct correspondence with those found by homogenization in equation~\eqref{eq:1fiber2d}. Hereafter, we work our way back from the properties of the cloak to the properties of the Willis-elastic lattice. Matching equation~\eqref{eq:cloakProperties} to equation~\eqref{eq:1fiber2d} permits to identify most of the lattice descriptors; the other descriptors are set based on an adopted discretization scheme. Note that the descriptors will be graded in space because the properties of the cloak are graded as well.

Start with a 1D lattice of parameter $r_o$ such that $r_o\t N$ is a lattice vector. The locations of the rigid bodies of the Willis-elastic lattice under construction are obtained by applying the conformal transformation $\gt\psi$ to the lattice points. Thus, the lattice vector of the Willis-elastic lattice is $r\t n$ with $r=\lambda r_o$, up to discretization error. The unit vector $\t n$ is given by $\t n=\t R\t N$. The nominal unit cell area is $A=r^2$; its actual value will ultimately depend on the spacing between fibers but only one fiber will be considered here. The mass $m=\rho A$ is deduced from $\rho = \rho_o$. The position of the center of mass $\t c$ is given by $\flip{\t c} = f \partial(\lambda\t n)$ for some arbitrary function $f$ and where, upon expansion,
\[
\partial(\lambda\t n) = \frac{1}{\lambda}\frac{\partial^2\t x}{\partial X_i \partial X_j}N_iN_j. 
\]
The remaining properties are the moment of inertia $J$, the spring constant $k=EA/r^2$, the misalignment $\theta$ and the matrix-body interaction stiffness $K$. These must solve the remaining design constraints. In particular,
\[
    \theta = \frac{\omega_o^2 f \rho_o}{2 E_o\lambda},\quad
    K = \omega_o^2 \frac{J}{A} + \frac{\omega_o^4 f^2 \rho_o^2}{2 E_o},\quad
    E = 2\lambda^2 E_o.
\]

This leaves certain freedom in choosing, say, $\theta$ and $K$. In the simulations presented in the next section, we lump $K$ and $J$ together, since they only appear in the combination $K-\omega_o^2J/A$. As for $f$, it is chosen so that $\theta$ is a fraction of $r$. Finally, with $\theta$ and $\t n$, the attachment vector $\t a$ is set to
\[
\t a = \frac{r}{2}\theta\cross\t n.
\]

This concludes the design of the cloak using a graded Willis-elastic mechanical lattice. Although the Willis behavior is valid throughout the subwavelength spectrum, the cloak is only operational at a single \emph{target frequency}, namely $\omega_o$, given that several of its parameters are dependent on frequency.

\subsection{Numerical demonstration}
Let $\{\t X\}$ be a domain in the 2D plane. For convenience, each position $\t X$ is identified with a complex number. This permits to define the conformal map $\gt\psi$ using the expression of a holomorphic function. Here, the adopted transformation
\[
\t x = \gt\psi(\t X) \equiv \frac12 \t X \left(1 + \sqrt{1 - 4a^2/\t X^2}\right)
\]
creates a semi-circular cavity of radius $a$ centered around the origin. Note that $\t x$ approaches $\t X$ as $\abs{\t X}\to \infty$ meaning that the transformation perturbs ever so slightly the northern, eastern and western far ends of the complex plane. Seen from these horizons, domains $\{\t X\}$ and $\{\t x\}$ are indistinguishable. To define the cloak's properties, the transformation gradient and Hessian are needed:
\[
\t F = \frac{\t x^2}{\t x^2 - a^2},\quad \t H = -2a^2\frac{\t x^3}{(\t x^2-a^2)^3}.
\]
In particular,
\[
\lambda = \sqrt{\t F\t F^\dagger} > 0,\quad 
\partial(\lambda\t n) = \frac{1}{\lambda}\t H\t N^2.
\]
\revise{Note that the adopted transformation, as well as $\t F$ and $\t H$, are singular at $\t X=\pm2a$, or equivalently $\t x=\pm a$. If needed, one can shift the transformation in space to ensure that these singularities do not occur within $\{\t X\}$.}

To assess the performance of the cloak, \revise{we investigate its ability to mimic one natural mode of the reference medium at one target frequency. First, the reference medium $\{\t X\}$ is discretized, its stiffness and mass matrices $\mathbb{K}_o$ and $\mathbb{M}_o$ are assembled and the eigenvalue problem
\[
(\mathbb{K}_o - \omega^2\mathbb{M}_o)\mathbb{U} = 0
\]
is solved. One eigenfrequency-eigenvector pair $(\omega_o,\mathbb{U}_o)$ is extracted and the frequency $\omega_o$ is set as the cloak's target frequency. The cloak on the other hand has been designed as a lattice and is therefore naturally discretized with stiffness matrix $\mathbb{K}(\omega_o)$ and mass matrix $\mathbb{M}(\omega_o)$. The dependence of these matrices on $\omega_o$ is made explicit to recall that the cloak's properties need be adjusted in function of the target frequency. Then, the eigenvalue problem
\[
(\mathbb{K}(\omega_o) - \omega^2\mathbb{M}(\omega_o))\mathbb{U} = 0
\]
is solved in the cloak and the eigenmode $(\omega,\mathbb{U})$ with the smallest discrepancy $\abs{\omega-\omega_o}$ is extracted. The cloak performs well if the differences between the eigenvectors $\mathbb{U}$ and $\mathbb{U}_o$, but also between the eigenfrequencies $\omega$ and $\omega_o$, are small. In fact, barring roundoff, discretization and convergence errors, the theory promises $\omega=\omega_o$ and $\mathbb{U}=\mathbb{U}_o$.}

\begin{figure}
    \centering
    \includegraphics[width=0.75\linewidth]{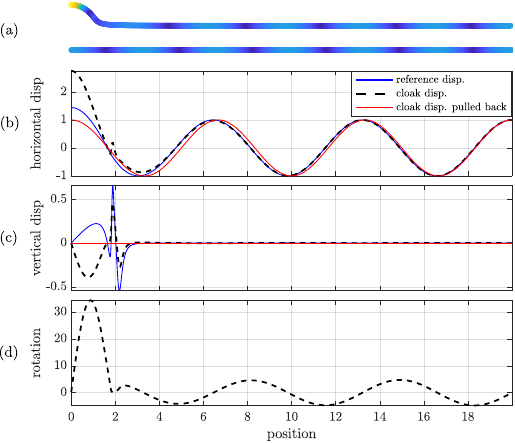}
    \caption{Displacement profiles plotted against position for an edge cavity: (a) contours of the horizontal displacements in the cloak (top) and the reference medium (bottom); (b) horizontal displacements; (c) vertical displacements; (d) rotations. Position is measured in units of $\t x$ in the cloak and in units of $\t X$ in the reference. Numerical parameters: cavity has $a=1$ radius; reference is $20$ units wide and is at $1/10$ above the $x$-axis; reference material properties $E_o=\rho_o=1$; discretization step $r_o=7.8\times10^{-3}$; target frequency $\omega_o=0.9425$.}
    \label{fig:fields1}
\end{figure}

\begin{figure}
    \centering
    \includegraphics[width=0.75\linewidth]{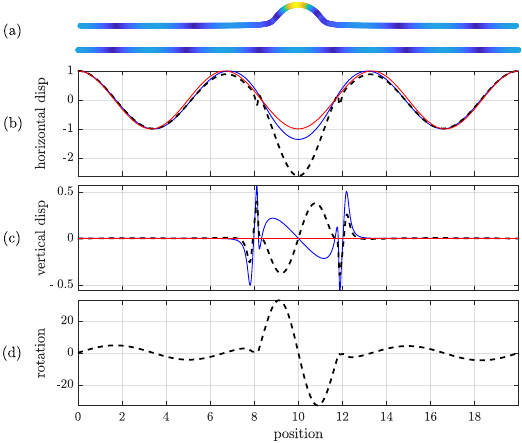}
    \caption{Displacement profiles plotted against position for a central cavity: (a) contours of the horizontal displacements in the cloak (top) and the reference medium (bottom); (b) horizontal displacements; (c) vertical displacements; (d) rotations. See Fig.~\ref{fig:fields1} for details.}
    \label{fig:fields2}
\end{figure}

\revise{The numerical results are summarized in Figs~\ref{fig:fields1} and~\ref{fig:fields2}. Two cases are analyzed: in Fig.~\ref{fig:fields1}, the cloak media has one free end near the cloaked cavity and one end far from it; in Fig.~\ref{fig:fields2}, both free ends extend far from the cloaked cavity.}  Three displacement profiles are plotted: one for the reference displacement $\t U$, one for the displacement in the cloak $\t u$ and one for the displacement in the cloak pulled-back into the reference medium, namely $\t F^\dagger\t u$. Displacements $\t U$ and $\t F^\dagger\t u$ should be identical, up to discretization error: horizontal components appear to match quite well; vertical components match well except at the points closest to the previously mentioned singularities at $\t X = \pm 2a$ where a boundary layer forms. On the other hand, the cloaking effect depends on $\t U$ and $\t u$ matching far from the cloaked cavity: this is observed in both cases and for both components under consideration. The profile of the normalized rotation $r \phi/\t u(\infty)$ in the cloak is also included to confirm the asymptotic scaling $\t u=O(r\phi)$; the scaling appears to hold but is degraded closest to the singularity at $\t x=\pm a$. Note that a wavelength comparable to the size of the cavity was chosen so as to favor scattering.

\begin{figure}
    \centering
    \includegraphics[width=0.7\linewidth]{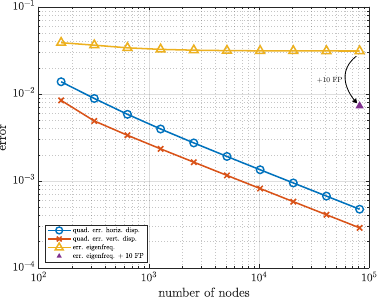}
    \caption{Convergence analysis: quadratic errors for horizontal and vertical components of reference and pulled-back displacements are plotted against the number of discretization points. Error on eigenfrequencies is included as well as one extra data point illustrating the benefit of the fixed-point scheme.}
    \label{fig:cv}
\end{figure}

A convergence analysis was further carried by decreasing the discretization step while maintaining constant all macroscopic properties. The results are summarized in Fig.~\ref{fig:cv}. The quadratic errors over the the horizontal and vertical components of the reference and pulled-back displacements $\t U$ and $\t F^\dagger\t u$ are plotted against the number of nodes. Both errors appear to decrease linearly with the step size. The error between the target frequency in the reference $\omega_o$ and the closest eigenfrequency in the cloak $\omega$ is also depicted but does not appear to decrease with step size. In other words, the eigenvectors match even if the eigenfrequencies do not. This is due to the fact that the eigenfrequency in the cloak $\omega$ depend on the cloak's material properties which themselves are $\omega_o$-dependent. To remedy this, the eigenfrequency in the cloak must be computed for converged material properties in the cloak. Thus, following a first computation of $\omega$ based on $\omega_o$, the properties in the cloak are corrected and repeatedly so in a fixed-point scheme of the form:
\[
    \omega_{i+1} \quad\text{minimizes}\quad \abs{\omega-\omega_i} \quad\text{among the eigenvalues of} \quad (\mathbb{K}(\omega_i),\mathbb{M}(\omega_i)).
\]
The fixed-point scheme significantly improves the convergence in terms of eigenfrequencies but has no significant influence over convergence for displacements. See the extra data point on Fig.~\ref{fig:cv}. Finally, note that, because of the boundary layer, convergence in infinity norm is not to be expected, not near the cloaked cavity in any case. Simulation code is available at \url{https://github.com/nassarh/trussx}.

\section{Conclusion}
The main finding of the paper is that Willis elasticity is a particular microcontinuum field theory with kinematic enrichment but where the enrichment has been eliminated in favor of the macroscopic displacement field through history-dependent, sometimes non-local boundary-condition dependent, convolution kernels. In particular, \emph{Willis elasticity models no novel phenomena that are not already within the reach of microcontinuum field theories}. That said, Willis elasticity proves useful in solving some inverse design problems. Here, a Willis-elastic invisibility cloak is designed, at the discrete micro-scale level, and is tested numerically. This is the first demonstration of cloaking of its kind where lattice-resolved Willis-elastic materials are used.

The above conclusion is somewhat disappointing. The silver lining is that it raises interesting questions regarding form-invariance under curvilinear changes of coordinates and transformation-based cloaking in microcontinuum field theories in general seeing how Willis elasticity turned out to be only a particular case of such theories.

\section*{Competing interests}
The authors declare no competing interests.

\section*{Acknowledgement}
Work supported by the NSF under CAREER award No. CMMI-2045881.


\begin{thebibliography}{10}
\expandafter\ifx\csname url\endcsname\relax
  \def\url#1{\texttt{#1}}\fi
\expandafter\ifx\csname urlprefix\endcsname\relax\def\urlprefix{URL }\fi
\expandafter\ifx\csname href\endcsname\relax
  \def\href#1#2{#2} \def\path#1{#1}\fi

\bibitem{Willis1980a}
J.~R. Willis, {A polarization approach to the scattering of elastic waves-I.
  Scattering by a single inclusion}, Journal of the Mechanics and Physics of
  Solids 28~(1970) (1980) 287--305.

\bibitem{Willis1980}
J.~R. Willis, {A polarization approach to the scattering of elastic waves-II.
  Multiple scattering}, Journal of the Mechanics and Physics of Solids
  28~(1976) (1980) 307--327.

\bibitem{Willis1981}
J.~R. Willis, {Variational principles for dynamic problems for inhomogeneous
  elastic media}, Wave Motion 3 (1981) 1--11.

\bibitem{Willis1985}
J.~R. Willis, {The nonlocal influence of density variations in a composite},
  International Journal of Solids and Structures 21~(7) (1985) 805--817.

\bibitem{Willis1997}
J.~R. Willis, {Dynamics of composites}, in: P.~Suquet (Ed.), Continuum
  Micromechanics, Springer-Verlag New York, Inc., 1997, pp. 265--290.

\bibitem{Willis2011}
J.~R. Willis, {Effective constitutive relations for waves in composites and
  metamaterials}, Proceedings of the Royal Society A 467~(2131) (2011)
  1865--1879.

\bibitem{Shuvalov2011}
A.~L. Shuvalov, A.~A. Kutsenko, A.~N. Norris, O.~Poncelet, {Effective Willis
  constitutive equations for periodically stratified anisotropic elastic
  media}, Proceedings of the Royal Society A 467~(2130) (2011) 1749--1769.

\bibitem{Nassar2015a}
H.~Nassar, Q.-C. He, N.~Auffray, {Willis elastodynamic homogenization theory
  revisited for periodic media}, Journal of the Mechanics and Physics of Solids
  77 (2015) 158--178.

\bibitem{Muhlestein2016}
M.~B. Muhlestein, C.~F. Sieck, A.~Al{\`{u}}, M.~R. Haberman, {Reciprocity,
  passivity and causality inWillis materials}, Proceedings of the Royal Society
  A: Mathematical, Physical and Engineering Science 472 (2016) 20160604.

\bibitem{Milton2006}
G.~W. Milton, M.~Briane, J.~R. Willis, {On cloaking for elasticity and physical
  equations with a transformation invariant form}, New Journal of Physics
  8~(10) (2006) 248--267.

\bibitem{Parnell2012}
W.~J. Parnell, A.~N. Norris, T.~Shearer, {Employing pre-stress to generate
  finite cloaks for antiplane elastic waves}, Applied Physics Letters 100
  (2012) 171907.

\bibitem{Parnell2012a}
W.~J. Parnell, {Nonlinear pre-stress for cloaking from antiplane elastic
  waves}, Proceedings of the Royal Society A 468 (2012) 563--580.

\bibitem{Norris2012a}
A.~N. Norris, W.~J. Parnell, {Hyperelastic cloaking theory: transformation
  elasticity with pre-stressed solids}, Proceedings of the Royal Society A 468
  (2012) 2881--2903.

\bibitem{Farhat2009}
M.~Farhat, S.~Guenneau, S.~Enoch, A.~B. Movchan, {Cloaking bending waves
  propagating in thin elastic plates}, Physical Review B - Condensed Matter and
  Materials Physics 79~(3) (2009) 033102.

\bibitem{Farhat2009a}
M.~Farhat, S.~Guenneau, S.~Enoch, {Ultrabroadband elastic cloaking in thin
  plates}, Physical Review Letters 103~(2) (2009) 024301.

\bibitem{Farhat2012}
M.~Farhat, S.~Guenneau, S.~Enoch, {Broadband cloaking of bending waves via
  homogenization of multiply perforated radially symmetric and isotropic thin
  elastic plates}, Physical Review B - Condensed Matter and Materials Physics
  85~(2) (2012) 020301.

\bibitem{Nassar2018a}
H.~Nassar, Y.~Chen, G.~L. Huang, {A degenerate polar lattice for cloaking in
  full two-dimensional elastodynamics and statics}, Proceedings of the Royal
  Society A 474 (2018) 20180523.

\bibitem{Nassar2019}
H.~Nassar, Y.~Y. Chen, G.~L. Huang, {Isotropic polar solids for conformal
  transformation elasticity and cloaking}, Journal of the Mechanics and Physics
  of Solids 129 (2019) 229--243.

\bibitem{Nassar2020}
H.~Nassar, Y.~Y. Chen, G.~L. Huang, {Polar metamaterials : A new outlook on
  resonance for cloaking applications}, Physical Review Letters 124~(8) (2020)
  84301.

\bibitem{Xu2020}
X.~Xu, C.~Wang, W.~Shou, Z.~Du, Y.~Chen, B.~Li, W.~Matusik, H.~Nassar, G.~L.
  Huang, {Physical realization of elastic cloaking with a polar material},
  Physical Review Letters 124~(11) (2020) 114301.

\bibitem{Chen2021}
Y.~Chen, H.~Nassar, G.~Huang, {Discrete transformation elasticity: An approach
  to design lattice-based polar metamaterials}, International Journal of
  Engineering Science 168~(March) (2021) 103562.

\bibitem{Nassar2017}
H.~Nassar, X.~C. Xu, A.~N. Norris, G.~L. Huang, {Modulated phononic crystals:
  Non-reciprocal wave propagation and Willis materials}, Journal of the
  Mechanics and Physics of Solids 101 (2017) 10--29.

\bibitem{Melnikov2019}
A.~Melnikov, Y.~K. Chiang, L.~Quan, S.~Oberst, A.~Al{\`{u}}, S.~Marburg,
  D.~Powell, {Acoustic meta-atom with experimentally verified maximum Willis
  coupling}, Nature Communications 10~(1) (2019) 1--7.

\bibitem{Milton2007a}
G.~W. Milton, {New metamaterials with macroscopic behavior outside that of
  continuum elastodynamics}, New Journal of Physics 9~(10) (2007) 359--359.

\bibitem{Boutin2018}
C.~Boutin, J.-L. Auriault, G.~Bonnet, {Inner Resonance in media governed by
  hyperbolic and parabolic dynamic equations. Principle and examples}, in:
  H.~Altenbach, J.~Pouget, M.~Rousseau, B.~Collet, T.~Michelitsch (Eds.),
  Generalized models and non-classical approaches in complex materials 1,
  Springer Nature, 2018, Ch.~6, pp. 83--134.

\bibitem{Qu2022}
H.~Qu, X.~Liu, G.~Hu, {Mass-spring model of elastic media with customizable
  Willis coupling}, International Journal of Mechanical Sciences 224~(5) (2022)
  107325.

\bibitem{Nassar2020c}
H.~Nassar, H.~Chen, G.~L. Huang, {Microtwist elasticity : A continuum approach
  to zero modes and topological polarization in Kagome lattices}, Journal of
  the Mechanics and Physics of Solids 144 (2020) 104107.

\bibitem{Mindlin1964}
R.~D. Mindlin, {Micro-structure in linear elasticity}, Archive for Rational
  Mechanics and Analysis 16 (1964) 51--78.

\bibitem{Germain1973}
P.~Germain, {The method of virtual power in continuum mechanics. Part 2:
  Microstructure}, SIAM Journal on Applied Mathematics 25~(3) (1973) 556--575.

\bibitem{CemalEringen1999}
A.~C. Eringen, {Microcontinuum field theories I: Foundations and solids},
  Springer, New York, 1999.

\bibitem{Spadoni2009}
A.~Spadoni, M.~Ruzzene, S.~Gonella, F.~Scarpa, {Phononic properties of
  hexagonal chiral lattices}, Wave Motion 46~(7) (2009) 435--450.

\bibitem{Bacigalupo2014}
A.~Bacigalupo, L.~Gambarotta, {Homogenization of periodic hexa- and tetrachiral
  cellular solids}, Composite Structures 116~(1) (2014) 461--476.

\bibitem{Rosi2016}
G.~Rosi, N.~Auffray, {Anisotropic and dispersive wave propagation within
  strain-gradient framework}, Wave Motion 63 (2016) 120--134.

\bibitem{Norris2008}
A.~N. Norris, {Acoustic cloaking theory}, Proceedings of the Royal Society A
  464~(2097) (2008) 2411--2434.

\bibitem{Norris2011}
A.~N. Norris, A.~L. Shuvalov, {Elastic cloaking theory}, Wave Motion 48~(6)
  (2011) 525--538.

\bibitem{Norris2015}
A.~N. Norris, {Acoustic cloaking}, Acoustics Today 11~(1) (2015) 38--46.

\bibitem{Milton2007}
G.~W. Milton, J.~R. Willis, {On modifications of Newton's second law and linear
  continuum elastodynamics}, Proceedings of the Royal Society A 463~(2079)
  (2007) 855--880.

\end{thebibliography}
\end{document}